\documentclass[11pt,a4paper]{article}
\usepackage[T1]{fontenc}
\usepackage[utf8]{inputenc}
\usepackage{mathptmx}
\usepackage[margin=1in]{geometry}
\usepackage{graphicx}
\usepackage{booktabs,tabularx,array}
\usepackage{microtype}
\usepackage[hidelinks]{hyperref}
\hypersetup{pdftitle={Richard: Voice-First Mobile Interaction for Persistent Tasks},
  pdfauthor={XinYang Chen}}
\title{Richard: Voice-First Mobile Interaction\\for Persistent Tasks}
\author{XinYang Chen}
\date{}
\begin{document}
\maketitle
\begin{abstract}
Mobile terminals need to provide application and network services while supporting users' control over their attention. We explore voice-first interaction organized around requests and delegated tasks, allowing users to leave a conversation and later inspect, revise, and retrieve the work. We present Richard, a system prototype that manages voice sessions, task execution, and result delivery separately, linking them through persistent request records. Conversation and task views provide visual feedback, while the backend coordinates immediate responses, dedicated service operations, and agent tasks. Request revisions, execution states, and notifications remain associated with the relevant task. We examine this design through Android functional records, controlled lifecycle verification, and execution records of a real programming request. Controlled verification reproduces revision, execution after confirmation, and result retention; deployed-service records show backend progress and failure feedback after client disconnection. These observations inform the design of task continuity, user control, and service integration in mobile voice interaction, providing an implementation basis for personal computing devices that accommodate intermittent user participation.
\end{abstract}


\section{Introduction}\label{sec:1}

Smartphone-centered mobile computing makes communication, information, and everyday services readily available, while concentrating these activities in an environment of competing demands on attention. In app-centric interaction, users typically enter interfaces, select operations, wait for feedback, and move between applications. Pursuing an initial goal can therefore expose users to unrelated content, notifications, and invitations to continue browsing. Research on digital self-control examines conflicts between intended and actual use, including frequent checking and spending longer on devices than intended.~\cite{lyngs2019} This raises a foundational design question: how can a mobile terminal provide computing capabilities while allowing users to direct their attention and disengage when they choose?

Prior work offers interventions for attention management and digital self-control. Lyngs et al. reviewed applications and browser extensions, identifying strategies including usage feedback and access restrictions.~\cite{lyngs2019} In a field experiment, Fitz et al. found that batching notifications three times daily improved participants' reported attentiveness and perceived control over phone use.~\cite{fitz2019} These findings show that interaction design can change attentional demands. We examine the participation required by task completion itself: can a terminal undertake operations and follow-up after a user expresses a goal, without requiring continued presence in interfaces or conversation? Calm technology's account of the center and periphery of attention provides a perspective on this direction.~\cite{weiser1996}

Richard accordingly explores a shift from operating applications to expressing requests, delegating execution, and inspecting work as needed. Its long-term objective is a portable terminal that progressively assumes everyday smartphone functions while allowing users to return their attention to activities beyond the device. Voice supports expression and clarification; visual feedback supports reading results and making complex comparisons. Reducing unnecessary continued participation also requires retaining tasks, managing subsequent progress, and providing clear opportunities for user decisions. Users determine their goals, participation, and continuation; the system organizes work around those decisions. Correction and intervention remain integral to this design.~\cite{amershi2019}

Natural-language interfaces and agent research already provide relevant capabilities. Iris composes operations through nested conversations,~\cite{fast2018} PUMICE combines language and demonstrations to clarify tasks,~\cite{li2019} AutoDroid operates Android interfaces,~\cite{wen2024} and Collaborative Gym supports human-agent interaction without fixed turn-taking.~\cite{shao2026} These systems establish foundations for delegation and asynchronous collaboration. The gap examined here concerns how these capabilities jointly support disengagement and re-engagement in a mobile terminal. When users stop operating interfaces or maintaining a conversation, what should persist, when should intervention be requested, and how can users later understand and control the same work?

This question requires separating user participation from task execution time. A user might request an expense-analysis program, end the conversation, later add a filtering requirement, and then inspect the delivered files. If execution depends on the current session, leaving can interrupt work. If execution continues without adequate feedback, users may need repeated checks to determine whether revisions took effect. Execution completion and result delivery can also occur at different times. Retained execution records, on-demand inspection, and distinct controls for ending conversation, muting speech, and cancelling tasks are concrete requirements for leaving and returning at will.

Richard implements these requirements by managing sessions, task execution, and result delivery separately, linking them through persistent request records. The paper contributes an interaction organization and rationale for intermittent participation, an architecture connecting a mobile client to an agent backend, and case analysis of execution across connections, request revision, and result retention. These mechanisms provide an implementation basis for organizing mobile computing around user goals and for subsequent research on attentional demands and autonomy over use.

\section{Related Work and Research Position}\label{sec:2}

\subsection{Expressing complex tasks and accessing services}\label{sec:2.1}

Natural-language interfaces translate user expressions into executable operations. Iris organizes Python functions into composable conversations and displays variables and intermediate results. PUMICE uses dialogue and demonstrations in third-party application interfaces to explain concepts and learn procedures. These systems show how complex task expression can work alongside visual feedback. Richard builds on this combination, concentrating on interaction after execution begins, when users leave, return, or revise their requests.~\cite{fast2018} ~\cite{li2019}

AutoDroid combines language models with application analysis to automate Android interface operations.~\cite{wen2024} Richard accommodates both dedicated tools and general agent execution behind a common task view. The execution path can depend on the available service capabilities. GUI automation is consequently a complementary execution method that can be combined with the task organization studied here.

\subsection{Asynchronous collaboration and mobile sessions}\label{sec:2.2}

Collaborative Gym provides shared task environments, bidirectional communication, and a notification protocol. Its web applications combine chat with shared workspaces, allowing users to act while an agent is working.~\cite{shao2026} This is an important precedent for asynchronous collaboration. Richard examines its implications for portable voice access: preserving execution after disconnection, recording when revisions take effect, and distinguishing execution completion, result playback, and user acknowledgement.

Table~\ref{tab:1} compares the design concerns discussed in the respective papers. It positions the research question; Richard's case evaluates its own implementation rather than providing a comparative performance benchmark.

\begin{table}[tbp]
\centering
\caption{Relationship between prior systems and Richard's design concerns.}
\label{tab:1}
\small
\renewcommand{\arraystretch}{1.18}
\begin{tabularx}{\linewidth}{@{}>{\raggedright\arraybackslash}p{.20\linewidth}>{\raggedright\arraybackslash}X>{\raggedright\arraybackslash}p{.31\linewidth}@{}}
\toprule
\textbf{Work} & \textbf{Related design} & \textbf{Richard's focus} \\
\midrule
Iris \cite{fast2018} & Nested conversations, function composition, and visual results & Managing delegated work across mobile sessions \\
PUMICE \cite{li2019} & Language and GUI demonstrations; retained concepts and procedures & Recording revisions and when they affect execution \\
AutoDroid \cite{wen2024} & Accessing Android application functions through GUI actions & Shared task state and user control across execution methods \\
Collaborative Gym \cite{shao2026} & Non-turn-taking collaboration, shared workspaces, and notifications & Relationships among mobile voice connections, execution, and delivery \\
\bottomrule
\end{tabularx}
\end{table}

ReAct is a representative approach to combining language-model reasoning with tool actions,~\cite{yao2023} and Hermes supplies the agent execution framework used in this prototype.~\cite{hermes} Richard studies the organization between the user and that framework: which states should persist, which controls should have distinct meanings, and how task outcomes should return to subsequent interactions.

\section{Interaction Design}\label{sec:3}

\subsection{Maintaining continuity through tasks}\label{sec:3.1}

The system retains the original request, relevant context, prior responses, and unresolved questions for delegated work. Subsequent interactions refer to these records through the task. A user can end a conversation and later query the same work, while the system can distinguish a revision from an independent request.

Controls act on explicit objects. Ending the conversation releases the voice connection while execution continues. Muting changes notification behavior while preserving execution. Cancellation requests termination of ongoing work; external writes that have already occurred require separate handling. These distinctions define button and voice-command semantics, avoiding a single implicit interpretation of ``stop'' across different processes.

Revisions during execution particularly require explicit feedback. Richard records the supplement while retaining the request version used by the current attempt. When that attempt ends, the task waits for confirmation before executing the updated requirements. This choice adds a confirmation step but distinguishes a recorded revision from a revision already applied to the current execution.

\subsection{Visual feedback for inspection and intervention}\label{sec:3.2}

\begin{figure}[tbp]
\centering
\includegraphics[width=.95\linewidth]{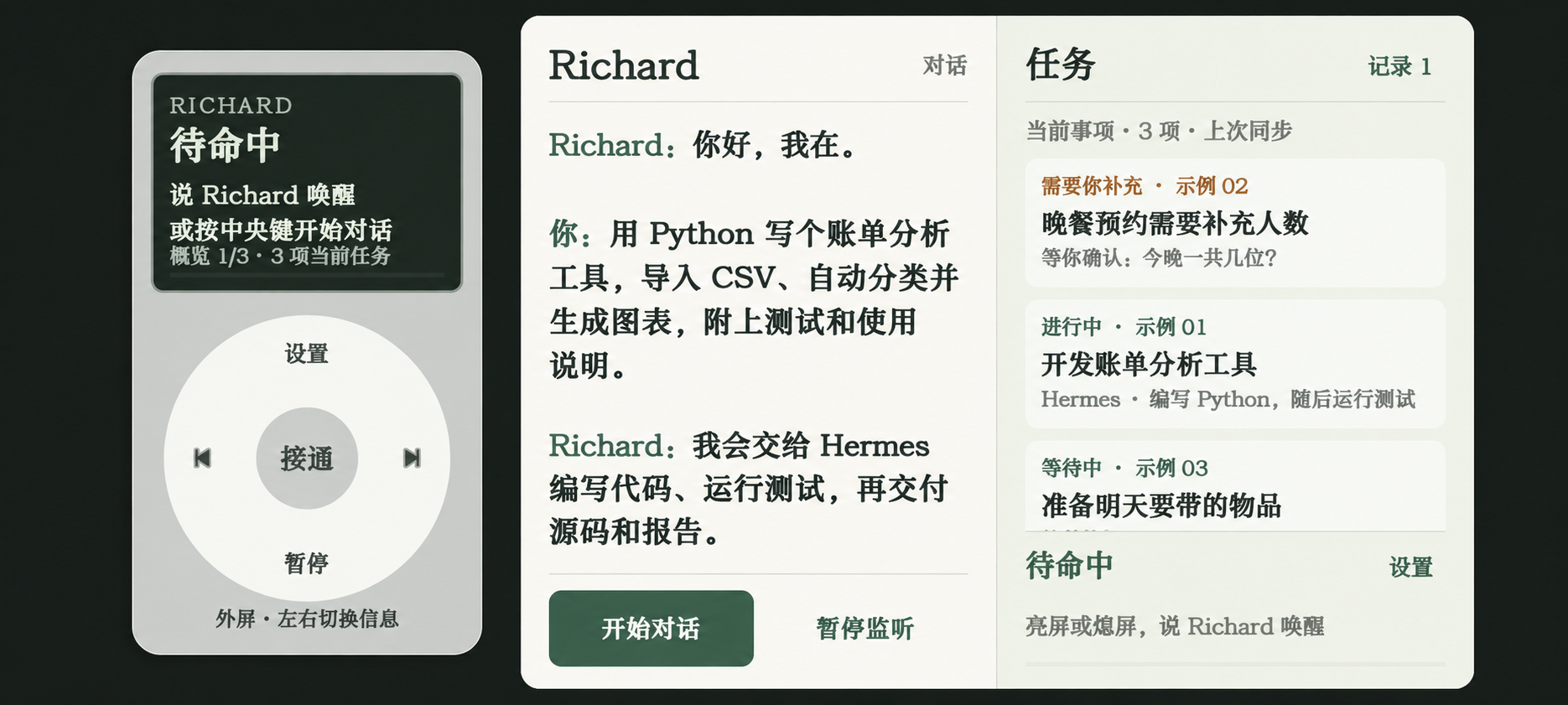}
\caption{Programming interaction illustration produced from an Android prototype screenshot using generative image editing. The simulated outer display is on the left; conversation and tasks occupy the two-page region on the right. The image explains the layout. Section~\ref{sec:5} separately examines execution records for a programming request.}
\label{fig:1}
\end{figure}

Voice supports expression and conversation, while the visual interface supports ongoing task inspection. In landscape orientation, the Android prototype places a two-page region with a $\sqrt{2}$:1 aspect ratio on the right: conversation on the left page and tasks with their states on the right page. A simulated outer display and circular controls occupy the remaining area, providing an overview and call controls. Both regions read the same task state. Figure~\ref{fig:1} illustrates the layout using a programming request.

This arrangement separates conversational history from current work, so progress can be inspected without reconstructing it from the entire dialogue. Routine tool activity updates the task view; errors, final outcomes, and requests for input can trigger speech notifications. On reconnection, the system first addresses the current interaction and retains historical results for explicit queries. The aspect ratio and simulated outer display are revisable prototype choices; the relationships among task state and session controls can be implemented on other form factors.

\section{System Architecture}\label{sec:4}

\subsection{From requests to execution paths}\label{sec:4.1}

\begin{figure}[tbp]
\centering
\includegraphics[width=1.0\linewidth]{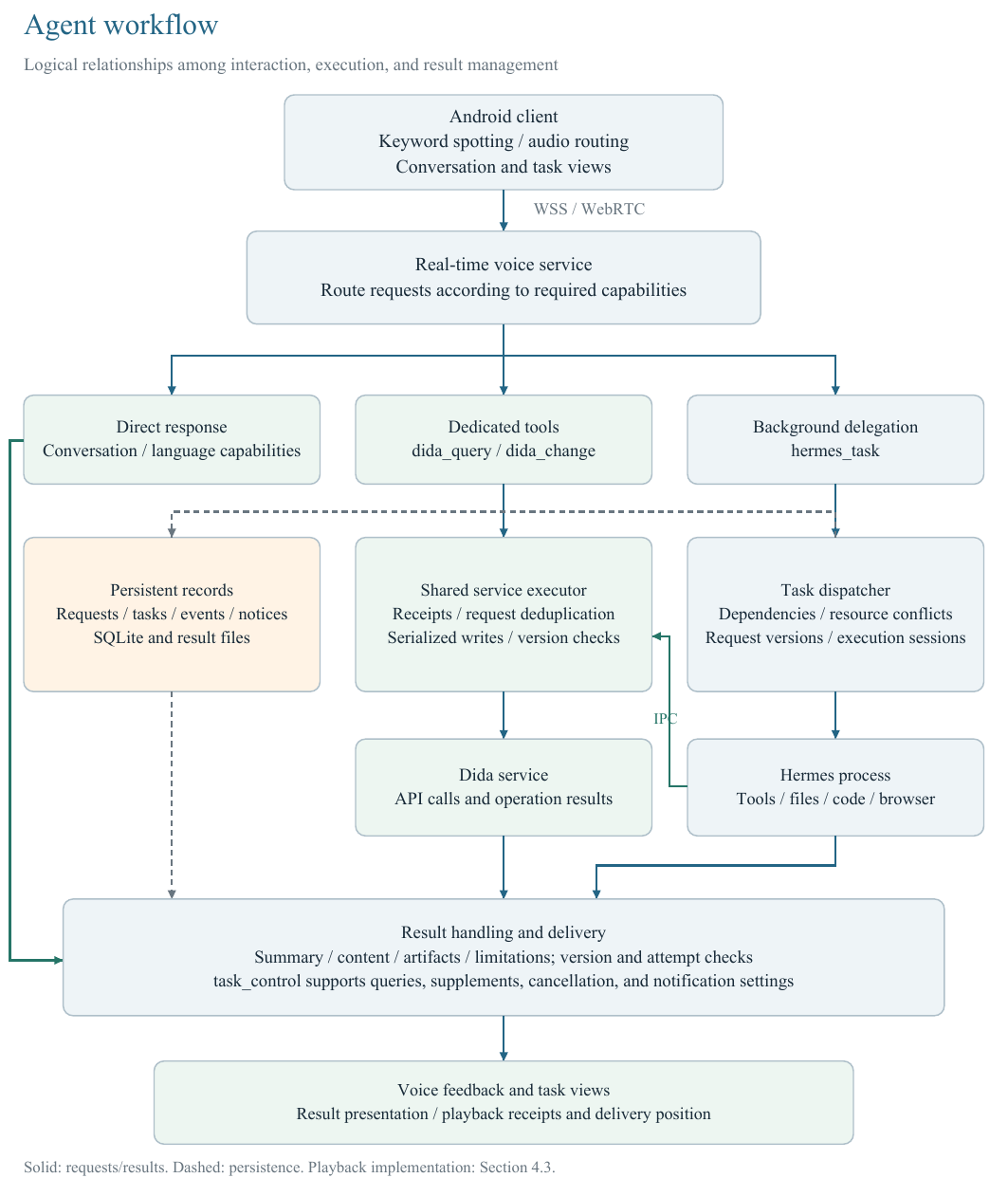}
\caption{Richard's agent workflow. Immediate responses, dedicated tools, and background tasks address different requests. Dedicated tools and Hermes share an executor when accessing the same service. Dashed connections represent persistence. Components are separated by logical responsibility and can share a deployment.}
\label{fig:2}
\end{figure}

Richard uses a real-time voice service for the current conversation and selects an execution path according to the capabilities needed. Explanations and language transformations can receive immediate responses. Routine Dida queries and changes use dedicated tools. Work involving files, code, or complex analysis is submitted to Hermes as a background task. Mixed requests can receive an initial grounded answer followed by information obtained through execution. Figure~\ref{fig:2} shows these paths and their shared state and result mechanisms.

\subsection{Independent execution and traceable revisions}\label{sec:4.2}

The task service returns an identifier when accepting a request and schedules execution in a process independent of the voice connection. A request can link multiple tasks; each task retains its execution attempt, workspace, and request version. The dispatcher checks dependencies and declared resource conflicts, recording completion, failure, and waiting for input separately. Progress queries read existing records directly, without starting another agent task.

\begin{figure}[tbp]
\centering
\includegraphics[width=1.0\linewidth]{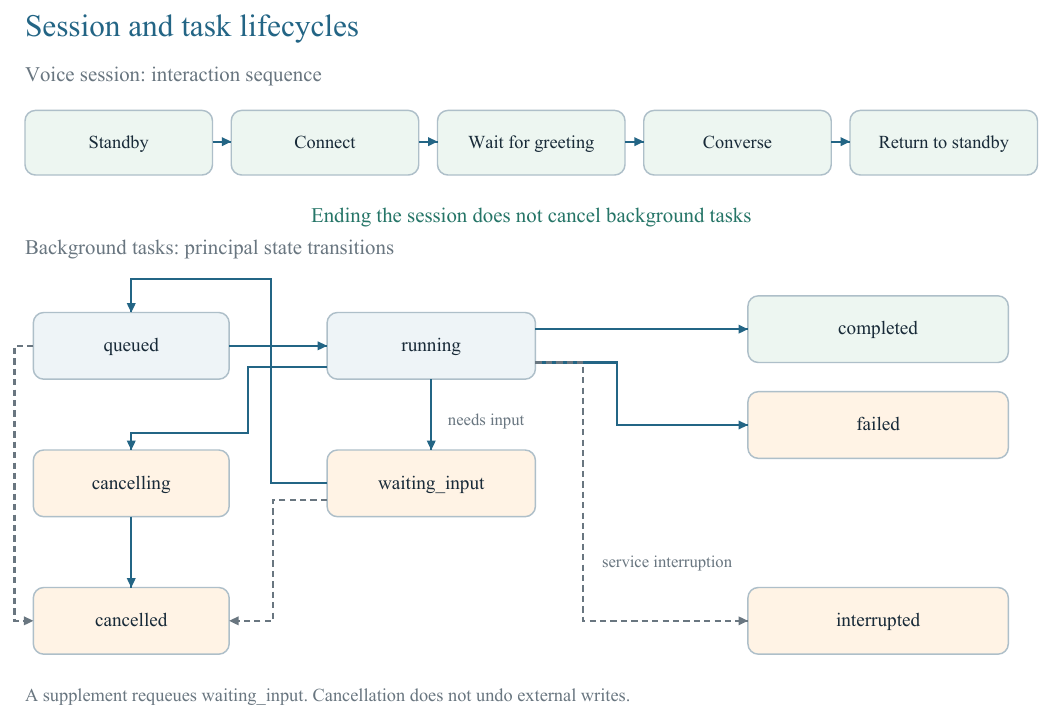}
\caption{Principal session and task transitions. A task continues after a connection ends. Supplementing a task that is waiting for input requeues it for another attempt. Dashed branches include service interruption and direct cancellation.}
\label{fig:3}
\end{figure}

Figure~\ref{fig:3} shows the principal session and task transitions. The request version is fixed when an attempt begins, and subsequent supplements are recorded separately. Once the result arrives, a version difference determines whether confirmation is needed. Results are therefore associated with both the requirements and the attempt that produced them. After a backend restart, previously running tasks are marked interrupted and retain their records for further action, avoiding automatic repetition of external writes.

\subsection{Recording delivery separately from completion}\label{sec:4.3}

Execution returns a summary, content, sources, and artifacts. The service stores results for later queries and paginated retrieval. Notifications are associated with a task, request version, and attempt; stale results are filtered and related events can be combined. Completion leaves the result available for inspection, with artifact quality assessed through the corresponding verification steps.

Speech delivery has a separate record. The backend advances the delivery position after a playback acknowledgement, retaining the task result when playback is interrupted. This path has been verified in the backend and web reference client; Android playback-acknowledgement integration remains follow-up work. Playback completion and user awareness represent a media event and a user acknowledgement, respectively, and are recorded separately.

\subsection{Sharing service operations across entry points}\label{sec:4.4}

Dida integration connects immediate service actions to complex background work. Voice tools and the Hermes local client access one executor. The executor reuses dida-cli's API module, stores operation receipts, deduplicates requests by identifier, and serializes writes to an account. A pre-submission version check detects changes that have already occurred; uncertain outcomes can be investigated through receipts.

Sharing the executor makes service rules independent of the calling entry point. Immediate requests and complex analysis tasks can follow different reasoning procedures while using consistent write and receipt handling. A race window remains between version checking and a cloud write. This mechanism makes operations traceable; stronger consistency depends on the transaction facilities of the external service.

\subsection{Prototype implementation}\label{sec:4.5}

The Kotlin client runs on Android 16 / API 36; ADB identifies the device as Motorola XT2533-4 / portov with an SM7435 SoC. CPU-based sherpa-onnx keyword spotting detects Richard,~\cite{sherpa} while standby input prioritizes a wired headset, a Bluetooth headset supporting call audio, and the built-in microphone. The client checks the actual recording route.~\cite{androidaudio} Following wake-up, it establishes the connection and users wait for the ready greeting before stating a request.

The backend runs in a container on a Jetson Nano 4GB. WSS and WebRTC carry signaling and media, while SQLite and files retain tasks and results. The current network deployment uses a relay, TURN, and an upstream proxy. The task service has five execution slots and a limit of twelve unfinished tasks. Workspaces and resource declarations coordinate concurrent access; operating-system isolation is a separate deployment concern. Detailed settings, source hashes, and case scripts accompany the manuscript.

\section{Feasibility and Case Analysis}\label{sec:5}

\subsection{Verification method}\label{sec:5.1}

Verification examines task continuity and user control. Mobile access is supported by functional records from the project initiator on one Android device. The programming case uses the deployed task interface and retains requests, state events, revisions, and error results. A script controls submission, disconnection, reconnection, and supplements to fix intervention timing, while Hermes invokes the actual model and tools. The case examines the task-service path; separate device records support mobile voice access.

\subsection{Controlled case: request revision and result retention}\label{sec:5.2}

We examined revision semantics using the actual task manager in the existing container, a new temporary database, and a deterministic Python subprocess. After a short wait, the subprocess returned fixed content identifying its execution attempt. This setup separated model-response uncertainty from state transitions, allowing request versions, attempts, and retrieved results to be checked directly.

The first attempt used request version 1. A supplement recorded during execution advanced the request to version 2, while the active process retained version 1. The task entered waiting\_input when the first attempt ended. Confirmation requeued the same task for a second attempt using version 3, which included the confirmation message. Reopening persistent storage after completion retained the completed state and the result ``attempt 2''; the delivery offset remained zero. Table~\ref{tab:2} summarizes the sequence. All corresponding assertions passed.

\begin{table}[tbp]
\centering
\caption{Request versions and execution state in the controlled case.}
\label{tab:2}
\small
\renewcommand{\arraystretch}{1.18}
\begin{tabularx}{\linewidth}{@{}>{\raggedright\arraybackslash}p{.20\linewidth}>{\raggedright\arraybackslash}X>{\raggedright\arraybackslash}p{.31\linewidth}@{}}
\toprule
\textbf{Action or event} & \textbf{Recorded requirements and execution basis} & \textbf{Task state} \\
\midrule
First attempt starts & Recorded version 1; executing version 1 & running, attempt 1 \\
Supplement during execution & Recorded version 2; still executing version 1 & running, attempt 1 \\
First attempt ends & Supplement has not yet been executed & waiting\_input \\
Confirmation and continuation & Recorded and executing version 3 & Attempt 2 completes \\
Storage reopened & Attempt 2 result retained; delivery offset zero & completed \\
\bottomrule
\end{tabularx}
\end{table}

The case verifies the correspondence among revision records, execution after confirmation, and retained results. A zero delivery offset at completion also illustrates that task completion and delivery are represented independently.

\subsection{Live case: a programming request and failure feedback}\label{sec:5.3}

We submitted a programming request to the deployed service: read five rows of synthetic expense CSV data, aggregate amounts by category, generate JSON and SVG outputs, and provide unit tests and usage instructions. The input included decimal amounts, a refund, and an empty category; the independently calculated acceptance total was 43.00. After execution started, the script added a month-filter requirement and closed the client connection. The backend continued, recording Hermes initialization and a terminal-tool invocation and return, showing that the connection ending did not terminate the task.

The first task, T027, checked its workspace and Python version, then received an upstream error reporting five consecutive attempts without a response and entered failed. The original request, supplement, execution version, and error result were retained. No program files were generated in the workspace. This execution therefore supplies a trace of progress and failure after disconnection; it supplies no artifact whose programming correctness can be checked.

A second independent task, T028, repeated the same request. It also remained waiting after the terminal check. After more than twelve minutes, the verification procedure requested cancellation, and the task reached cancelled without generating files. This externally stopped run is retained separately from T027's upstream-error termination.

This observation distinguishes continuity from successful execution. Connection independence allows work to continue, while an external model service can still prevent progress. A terminal needs to retain requirements and execution context on failure so that the user can decide what to do next. Neither task acceptance nor a successful tool invocation establishes fulfillment of the request. The request, complete event sequence, and independently prepared artifact acceptance script accompany the manuscript.

\subsection{Mobile access and external services}\label{sec:5.4}

The built-in microphone, a wired headset, and HUAWEI FreeClip each underwent actual input-route checks, with the tester confirming wake-up and the ready greeting. Screen-off wake-up was also observed with the built-in microphone. Client version 0.2.6 restored standby after an actual reboot observed over USB, followed by successful screen-off wake-up. With prior authorization, boot recovery opens a visible activity before starting the microphone foreground service to meet Android startup restrictions; configured lock-screen credentials also require an initial unlock.~\cite{androidforeground}

Existing service verification includes a real voice-service/Hermes round trip: the service first answered a general question, Hermes read a controlled file, and the actual cause and checkword returned for synthesis by the voice service. Dida verification covered temporary-task creation, reuse of the original receipt on a duplicate request, update readback, and deletion. A real Hermes task also verified query receipts through the shared executor. These records support integrating different execution paths into common task management; conditions and evidence locations are listed in the accompanying material.

\section{Discussion}\label{sec:6}

\subsection{System requirements for intermittent participation}\label{sec:6.1}

The live request records and controlled verification illustrate how stopping an interaction can be separated from stopping work. Returning to the interaction requires more state than simply reconnecting to a model. Task identity preserves the work item, request versions identify execution requirements, and delivery records describe whether results have been presented. Together, these records allow users to inspect progress during a later interaction and choose to continue, revise, or cancel.

This design relates to established asynchronous collaboration. Collaborative Gym coordinates activity in shared environments,~\cite{shao2026} while Richard includes voice connections and result delivery in mobile interaction management. A reusable design implication is to partition state and controls according to the meaning of user actions: connection state describes whether interaction is available, task state describes execution progress, and delivery state describes result presentation. These distinctions jointly inform the interface and notification behavior.

Applying an in-flight revision in a later attempt trades responsiveness for explicitness. Revising an active plan might be faster, but it requires explaining which operations have already occurred and which will adopt the new requirements. The current prototype uses attempt boundaries to make this inspectable. More fine-grained intervention should retain the correspondence between revisions and actual execution.

\subsection{Visual feedback and experience evaluation}\label{sec:6.2}

Presenting conversation alongside tasks supports inspection while retaining voice access. Because speech is sequential, complex comparisons and reading still benefit from visual feedback. Subsequent experience studies should examine whether users correctly interpret task states, recognize when intervention is required, and experience a different attention burden when fewer interface operations are needed.

The evidence consists of single-device use, controlled lifecycle verification, and live request records, supporting executability in the tested configuration. Diversity of natural requests, extended portable use, and the suitability of notification policies require subsequent use studies. A subsequent programming case should independently verify actual artifacts and examine the complete experience of using the system through mobile speech.

\subsection{Tools, skills, and third-party extension}\label{sec:6.3}

Dida integration provides a starting point for incremental third-party access. A command-line interface (CLI) or API reuses service functions; tool interfaces specify callable operations, parameters, and results; skills retain reusable procedures and operating instructions. Model Context Protocol (MCP) offers an option for tool discovery and invocation.~\cite{mcp} These components serve different roles and can be combined according to service requirements.

Third-party developers could consequently supply capabilities and procedures while the terminal retains common conversation, task, and result views. Further work should define interface descriptions, version compatibility, authorization scope, and error representation, ensuring that new entry points reuse operation receipts and write rules. Programs enforce permissions, while skills guide procedures. The current Dida voice path uses the shared executor; MCP access is a direction for extending it.

\subsection{Longer-term terminal development}\label{sec:6.4}

Longer-term research proceeds through dedicated hardware, visual perception, and an open personal computing system. The first stage starts with an ESP32 engineering prototype to investigate portable audio, display, power, and mechanical design, with a camera interface reserved. Low-power keyword detection requires further work on DSP model and driver compatibility. The second stage introduces camera or shared-screen input for task interpretation and result inspection. The third explores rules for integrating third-party tools and skills across everyday activities.

Across these stages, the design retains user requests and execution state while offering explicit queries and interventions. Processors, screen proportions, input devices, and agent frameworks can change. The phone prototype allows these interaction relationships to be examined before they guide dedicated hardware choices.

\section{Conclusion}\label{sec:7}

Richard investigates voice-first mobile interaction with intermittent user participation. It separates sessions, task execution, and result delivery while linking them through persistent request records. The prototype and cases illustrate execution across connections, request revision, result retention, and failure records. The work provides reusable design guidance for voice-first terminals: organize work through persistent tasks, support inspection through visual feedback, and preserve intervention through controls with explicit targets.

\clearpage

{\small



}


\begin{thebibliography}{99}

\bibitem{lyngs2019}
Ulrik Lyngs, Kai Lukoff, Petr Slovak, Reuben Binns, Adam Slack, Michael Inzlicht, Max Van Kleek, and Nigel Shadbolt. Self-Control in Cyberspace: Applying Dual Systems Theory to a Review of Digital Self-Control Tools. CHI 2019. DOI: \href{https://doi.org/10.1145/3290605.3300361}{10.1145/3290605.3300361}. \href{https://arxiv.org/abs/1902.00157}{Author version}.

\bibitem{fitz2019}
Nicholas Fitz, Kostadin Kushlev, Ranjan Jagannathan, Terrel Lewis, Devang Paliwal, and Dan Ariely. Batching smartphone notifications can improve well-being. Computers in Human Behavior 101 (2019), 84-94. DOI: \href{https://doi.org/10.1016/j.chb.2019.07.016}{10.1016/j.chb.2019.07.016}.

\bibitem{weiser1996}
Mark Weiser and John Seely Brown. The Coming Age of Calm Technology. 1996. \href{https://calmtech.com/papers/coming-age-calm-technology}{Reprinted original}.

\bibitem{amershi2019}
Saleema Amershi et al. Guidelines for Human-AI Interaction. CHI 2019. DOI: \href{https://doi.org/10.1145/3290605.3300233}{10.1145/3290605.3300233}.

\bibitem{fast2018}
Ethan Fast, Binbin Chen, Julia Mendelsohn, Jonathan Bassen, and Michael S. Bernstein. Iris: A Conversational Agent for Complex Tasks. CHI 2018. DOI: \href{https://doi.org/10.1145/3173574.3174047}{10.1145/3173574.3174047}.

\bibitem{li2019}
Toby Jia-Jun Li, Marissa Radensky, Justin Jia, Kirielle Singarajah, Tom M. Mitchell, and Brad A. Myers. PUMICE: A Multi-Modal Agent that Learns Concepts and Conditionals from Natural Language and Demonstrations. UIST 2019. DOI: \href{https://doi.org/10.1145/3332165.3347899}{10.1145/3332165.3347899}.

\bibitem{wen2024}
Hao Wen et al. AutoDroid: LLM-powered Task Automation in Android. MobiCom 2024. DOI: \href{https://doi.org/10.1145/3636534.3649379}{10.1145/3636534.3649379}. \href{https://arxiv.org/abs/2308.15272v4}{Author version}.

\bibitem{shao2026}
Yijia Shao, Vinay Samuel, Yucheng Jiang, John Yang, and Diyi Yang. Collaborative Gym: A Framework for Enabling and Evaluating Human-Agent Collaboration. ICLR 2026. \href{https://arxiv.org/abs/2412.15701v6}{arXiv:2412.15701v6}.

\bibitem{yao2023}
Shunyu Yao et al. ReAct: Synergizing Reasoning and Acting in Language Models. ICLR 2023. \href{https://arxiv.org/abs/2210.03629}{arXiv:2210.03629}.

\bibitem{hermes}
Nous Research. \href{https://github.com/NousResearch/hermes-agent}{Hermes Agent}. Project source and documentation. Accessed 2026-09-29.

\bibitem{sherpa}
k2-fsa contributors. \href{https://k2-fsa.github.io/sherpa/onnx/kws/index.html}{sherpa-onnx: Keyword spotting}. Accessed 2026-09-29.

\bibitem{androidaudio}
Android Developers. \href{https://developer.android.com/reference/android/media/AudioRecord\#getRoutedDevice()}{AudioRecord: getRoutedDevice}. Accessed 2026-09-29.

\bibitem{androidforeground}
Android Developers. \href{https://developer.android.com/about/versions/15/changes/foreground-service-types}{Changes to foreground service types for Android 15}. Accessed 2026-09-29.

\bibitem{mcp}
Model Context Protocol. \href{https://modelcontextprotocol.io/docs/learn/architecture}{Architecture overview}. Official documentation. Accessed 2026-09-29.

\end{thebibliography}
\end{document}